\documentclass[aps,prl,twocolumn,showpacs,superscriptaddress,preprintnumbers]{revtex4}
\usepackage{tabularx}
\usepackage{bm}
\usepackage{subfigure}
\usepackage{euscript}
\usepackage{epsfig,psfrag,subfigure}
\usepackage{graphicx,psfrag,subfigure}
\usepackage{color}
\usepackage{amsmath,amsfonts}
\usepackage{exscale}
\usepackage{amsbsy}
\usepackage{subfigure}

\def\be{\begin{equation}}
\def\ee{\end{equation}}
\def\ba{\begin{eqnarray}}
\def\ea{\end{eqnarray}}

\def\rr{{\bf r}}

\renewcommand\d{\partial}
\newcommand\drm{\mathrm{d}}
\newcommand\p{\mathbf{p}}

\begin{document}
\preprint{INT-PUB-12-029}
\title{Chiral Anomaly and Classical Negative Magnetoresistance
of Weyl Metals}
\author {D.~T.~Son}
\affiliation{Institute for Nuclear Theory, University of Washington,
Seattle, WA 98195-1550, USA}
\author {B.~Z.~Spivak}
\affiliation{Department of Physics, University of Washington,
Seattle, WA 98195, USA}

\date{June 2012}

\begin{abstract}
We consider the classical magnetoresistance of a Weyl metal in which
the electron Fermi surface possess nonzero fluxes of the Berry curvature.
Such a system may exhibit large negative magneto-resistance
with unusual anisotropy as a function of the angle between the
electric and magnetic fields.  In this case the system can support a 
new type of plasma waves. These phenomena are consequences of
chiral anomaly in electron transport theory.
\end{abstract}

\maketitle

Materials with nontrivial topological properties have attracted
considerable interest after the discovery of topological
insulators~\cite{TI}.  One type of such materials is the so-called
Weyl semimetals, characterized by the the presence of points of band
touching (Dirac
points)~\cite{Vishwanath,Balents,Balents1,GangXu,Balcond,Sid,Ran,Bernevig,
  Qi}.  In this paper, we study the metallic counterparts of these
materials---the Weyl metals, where Dirac points are hidden inside a
Fermi surface.  We show that these materials may exhibit large
negative magnetoresistance with unusual anisotropy.  We also find a
new type of plasma waves in these systems.

At low magnetic field ${\bf B}$ and at relatively high temperature
Landau quantization can be neglected, and electron transport
in metals can be described using the
semiclassical Boltzmann kinetic equation
\begin{equation}\label{kineq}
  \frac{\partial n_{{\bf p}}}{\partial t}
  + \dot{\bf r}\cdot\frac{\partial n_{{\bf p}}}{\partial {\bf r}}
    + \dot{\bf p}\cdot\frac{\partial n_{{\bf p}}}{\partial {\bf p}}
  = I_{\rm coll} \{n_{{\bf p}}\}.
\end{equation}
Here $n_{{\bf p}}({\bf r},t)$ is the electron distribution function,
${\bf p}$ is the quasimomentum, $I_{\rm coll} \{n_{{\bf p}}\}$
is the collision integral,  and
\begin{subequations}\label{eqmotion}
\begin{align}
  \dot \rr &= \frac{\partial \epsilon_{{\bf p}}}{\partial {\bf p}}
    + \dot\p\times\bm{\Omega}_{{\bf p}},\label{eqmotion-1} \\
  \dot \p &= e\mathbf{E} + \frac ec \dot\rr \times \mathbf{B}.
\end{align}
\end{subequations}
The last, ``anomalous,'' term in Eq.~(\ref{eqmotion-1}), proportional to the
Berry curvature
\begin{equation}
\bm{\Omega}_{{\bf p}}
 =\bm{\nabla}_{{\bf p}} \times {\bf A}_{\bf p},  \qquad
 {\bf A}_{\bf p}=i\langle  u_{{\bf p}}|\nabla_{{\bf p}}u_{{\bf p}}\rangle ,
\end{equation}
was introduced in Ref.~\cite{Niu}. (See also reviews on the subject in
Refs.~\cite{Revs,Ong}.)  In  systems with time-reversal symmetry
$\bm{\Omega}_{{\bf p}}=\bm{\Omega}_{-{\bf p}}$, while in
centro-symmetric systems $\bm{\Omega}_{{\bf p}}=-\bm{\Omega}_{-{\bf p}}$.
Thus, in systems which are both time- and centro-symmetric
${\bf \Omega}_{{\bf p}}=0$.  In this case the magneto-resistance
described by Eq.~(\ref{kineq}) is positive and is governed by
the parameter $(\omega_{\rm c}\tau_{\textrm{tr}})^{2}$~\cite{Abrikosov}.
Here $\omega_{\rm c}$ is the cyclotron frequency and $\tau_{\textrm{tr}}$
is the electron transport mean free path.
The Berry curvature is divergence-free except at isolated points in
${\bf p}$ space, which are associated with band degeneracies.
As a result, in the case where the electronic spectrum has several valleys,
they can be characterized by integers (see, for example, Ref.~\cite{Haldane})
  \begin{equation}
  k^{(i)} = \frac1{2\pi\hbar} \oint\!\drm\mathbf{S}\cdot
  \bm{\Omega}^{(i)}_{{\bf p}}=0, \pm 1, ....
\end{equation}
Here the index $"i"$ labels the valleys, $\drm\mathbf{S}$ is the elementary
area vector.  Nonzero values of
$k^{(i)}$ are realized if
near the degeneracy  points, electrons can be described by the massless Dirac
Hamiltonian~\footnote{In this paper, we use ``Dirac'' and ``Weyl''
interchangeably.}
\begin{equation}\label{DiracEq}
 H=\pm v\bm{\sigma}\cdot\hat{{\bf P}}.
\end{equation}
Here $\hat{{\bf P}}= -\hbar \bm{\nabla}-\frac{e}{c}{\bf A}$ is the momentum
operator, ${\bf A}$ is the vector potential, $\bm{\sigma}$ is the operator
of pseudospin, $v$ is the quasiparticle velocity, and the signs $\pm$
correspond to the different chiralities of the Weyl fermions.

It is well known that massless Dirac fermions exhibit chiral anomaly
which can be understood in the language of level crossing
in the presence of a magnetic field~\cite{Nielsen}.
According to the Nielsen-Ninomiya theorem~\cite{theorem},
the number of valleys with opposite chiralities (positive and negative
values of $k^{(i)}$) should be equal, and so $\sum_{i}k^{(i)}=0$.
Recently, gapless semiconductors with topologically protected
Dirac points (Weyl semi-metals) have attracted  significant
attention~\cite{Vishwanath,Balents,Balents1,GangXu,Balcond,Sid,Ran,Bernevig,
Qi}.
Both time-reversal-breaking~\cite{Vishwanath}, and
non-centro-symmetric~\cite{Balents1} versions of these systems have been
proposed.
In the absence of a random potential and doping, the chemical potentials
in these systems are at the Dirac points.

In this article we consider the case where in equilibrium the chemical
potential $\mu=\mu_{i}$ measured from the Dirac points is finite, and
show that the semiclassical Eqs.~(\ref{kineq}) and (\ref{eqmotion})
can yield a substantial anomaly-related negative magnetoresistance.
The latter also exhibits unusual anisotropy as a function of angle
$\theta$ between ${\bf E}$, and ${\bf B}$. Here $\mu_{i}$ is the chemical
potential in the $i$-th valley, measured from the Weyl's point.
Using Eqs.~(\ref{eqmotion}) we get
\begin{align}\label{eqmotion1}
  \dot \rr &= \left(1+\frac ec \mathbf{B}\cdot\bm{\Omega}_{\bf p}\right)^{-1}
  \left[ \mathbf{v} + e\mathbf{E}\times\bm{\Omega}_{\bf p} + \frac ec
  (\bm{\Omega}_{\bf p}\cdot\mathbf{v})\mathbf{B}\right],\nonumber \\
  \dot \p &= \left(1+\frac ec \mathbf{B}\cdot\bm{\Omega}_{\bf p}\right)^{-1}
  \left[ e\mathbf{E}+\frac ec\mathbf{v}\times\mathbf{B}
  + \frac{e^2}c (\mathbf{E}\cdot\mathbf{B})\bm{\Omega}_{\bf p} \right],
\end{align}
where $\mathbf{v}=\partial\epsilon_\p/\partial\p$.
Substituting Eqs.~(\ref{eqmotion1}) into Eq.~(\ref{kineq}) we get the
kinetic equation in the form
\begin{widetext}
\begin{multline}\label{kineq1}
  \frac{\partial n^{(i)}_{\bf p}}{\partial t}
  + \left( 1+ \frac ec \mathbf{B}\cdot\bm{\Omega}^{(i)}_{\bf p}\right)^{-1}
  \left[\left( e{\bf E}+\frac ec{\bf v}\times{\bf B}
  + \frac{e^2}c({\bf E \cdot B}) {\bf \Omega}^{(i)}_{{\bf p}}\right)
    \frac{\partial n^{(i)}_{{\bf p}}}{\partial {\bf p}} \right.\\
  \left. +
  \left( \mathbf{v} + e\mathbf{E}\times\bm{\Omega}^{(i)}_{\bf p} + \frac ec
  (\bm{\Omega}^{(i)}_{\bf p}\cdot\mathbf{v})\mathbf{B} \right)
  \frac{\partial n^{(i)}_{{\bf p}}}{\partial {\bf r}}\right] =I^{(i)}_{\textrm{coll}}\{n^{(i)}_{{\bf p}}\}
\end{multline}
\end{widetext}
(cf.\ Ref.~\cite{Duval:2005vn}).
Let us consider the case where $\mu\gg T$, $
\hbar \omega_{c}=\hbar |e|v^{2}B/c\mu$, and assume that the conductivity
of the system is determined by elastic scattering.
Then, the collision integral $I_\textrm{coll}$ in Eq.~(\ref{kineq1})
describes  elastic intra- and inter-valley scattering.
We assume that $\tau_{\textrm{tr}}\ll \tau $, where  $\tau$ is an elastic
inter-valley scattering mean free time. In this case the anisotropy of
the intra-valley distribution function can be neglected,
and the latter depends only on the energy $\epsilon$:
$n^{(i)}_\p=n^{(i)}(\epsilon)$.
Denoting by $\rho^{(i)}(\epsilon)$ the density of states~\cite{Revs,Ong},
\begin{equation}
  \rho^{(i)}(\epsilon) = \int\!\frac{\drm\p}{(2\pi\hbar)^3} \left(
    1 + \frac ec\mathbf{B}\cdot\bm{\Omega}^{(i)}_{{\bf p}}\right)
   \delta(\epsilon_\p-\epsilon),
\end{equation}
in the homogeneous case we get the kinetic equation in a form
\begin{equation}\label{kin-eps}
 \frac{\d n^{(i)}(\epsilon)}{\d t}
   + \frac{k^{(i)}}{\rho^{(i)}(\epsilon)} \frac{e^2}{4\pi^2\hbar^2c}
   (\mathbf{E}\cdot\mathbf{B})
   \frac{\d n^{(i)}(\epsilon)}{\d\epsilon}
   =I^{(i)}_{\textrm{coll}}\{n^{(i)}(\epsilon)\},
\end{equation}
where the collision integral now includes only inter-valley scattering.
For this, we will use the relaxation time approximation
\begin{equation}
  I^{(i)}_{\textrm{coll}}= -\frac{\delta n^{(i)}(\epsilon)}{\tau},
\end{equation}
where $\delta n^{i}(\epsilon)$ is the deviation of the distribution
function from its equilibrium value.

The electron density and entropy density in $i$-th valley are
\begin{align}
  N^{(i)} & = \int\!\drm\epsilon\,\rho^{(i)}(\epsilon)n^{(i)}(\epsilon)\\
   S^{(i)} &= -\!\int\! \drm\epsilon\, \rho^{(i)}(\epsilon),
  \left[ (n^{(i)}(\epsilon) \ln n^{(i)}(\epsilon) \right.\nonumber\\
   & \qquad\qquad\left.
   +(1-n^{(i)}(\epsilon))\ln (1-n^{(i)}(\epsilon))\right].
\end{align}
Integrating Eq.~(\ref{kin-eps}) over $\rho^{(i)}(\epsilon)\drm\epsilon$ we
get the conservation law for particle number in each valley:
\begin{equation}\label{chiralnonconserv}
  \frac{\d N^{(i)}}{\d t} + \bm{\nabla}\cdot \mathbf{j}^{(i)}
   = k^{(i)}\frac{e^2}{4\pi^2\hbar^2c}
    (\mathbf{E}\cdot\mathbf{B}) -\frac{\delta N^{(i)}}\tau\,,
\end{equation}
\begin{equation}\label{current}
  \mathbf{j}^{(i)}= \int\!\frac{\drm\p}{(2\pi\hbar)^3} \left[
  {\bf v}+ e\mathbf{E}\times \bm{\Omega}^{(i)}_{\bf p}
  +
   \frac ec
  \bigl(\bm{\Omega}^{(i)}_{{\bf p}}\cdot {\bf v}\bigr)\mathbf{B}
  \right]n_{{\bf p}}^{(i)}.
\end{equation}
Thus, in the presence of electric and magnetic field, the number of
particles in the $i$-th valley, $N^{(i)}$,
is not conserved even if $\tau \rightarrow \infty$.  This is the chiral
anomaly which was originally introduced  in field theory in Refs.~\cite{ADB},
and latter  discussed in the context of  electron band structure theory in
Ref.~\cite{Nielsen}, and in the theory of superfluid
$^3$He~\cite{volovik,Stone}.  It is interesting that the
anomaly can be understood completely in the framework of semiclassical
kinetic equation Eq.~(\ref{kineq}), characterized by
$k^{(i)}$~\cite{Son:2012wh}, and that the term proportional to
${\bf E}\cdot{\bf B}$ in Eq.~(\ref{chiralnonconserv}) is the same as
obtained in Ref.~\cite{Nielsen} in the ultra-quantum limit.

\begin{figure}[ptb]
\includegraphics[width=0.45\textwidth]{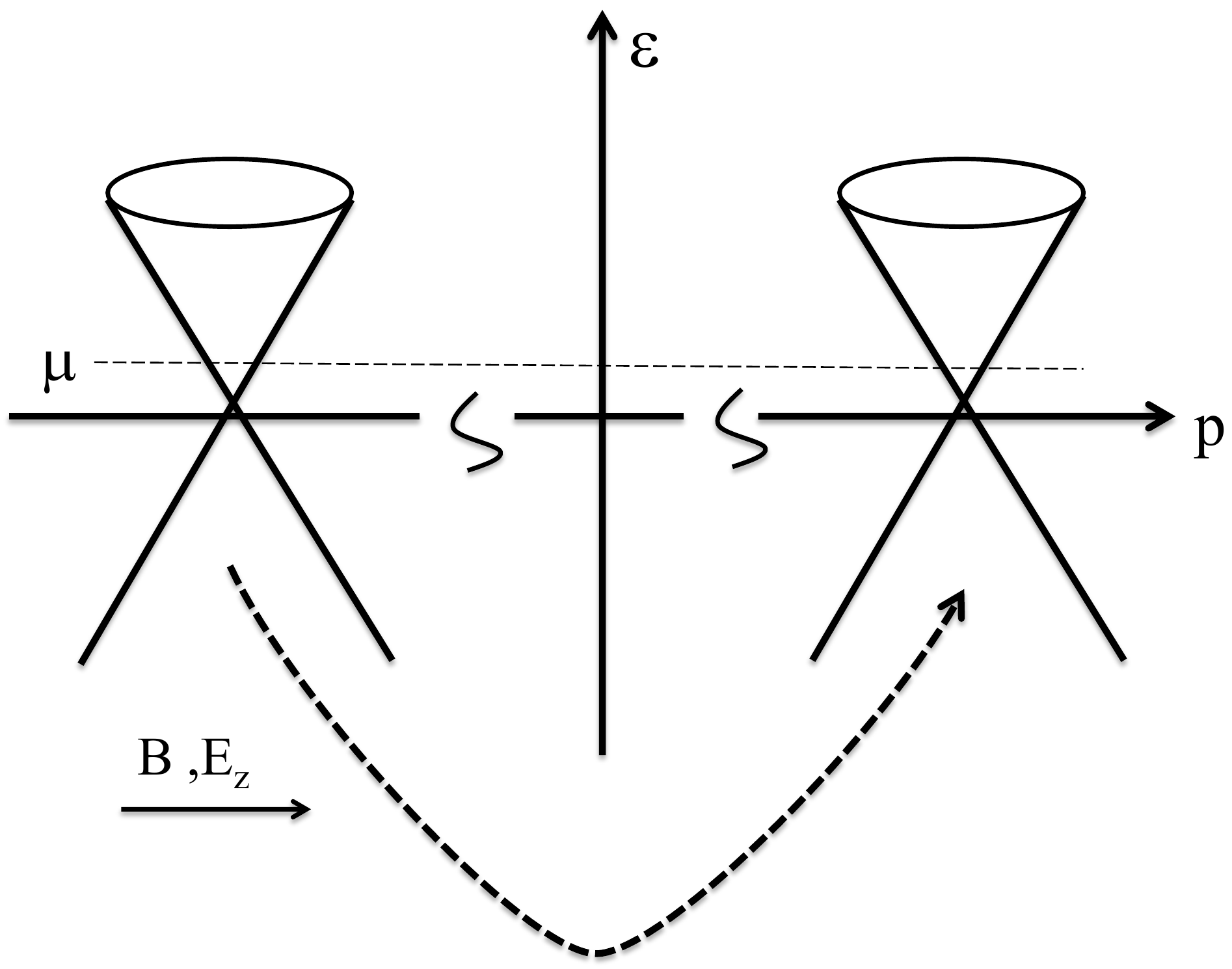}
\caption{Schematic 3D electron spectrum in a Weyl metal. Only two valleys
in the electron spectrum are shown. The  dashed line indicates the
direction of the electron spectral flow in the presence of parallel
electric and magnetic fields.} \label{fig:1}
\end{figure}

Equation~(\ref{kineq1}) represents a low energy effective theory. To
see why the number of electrons in an individual valley is not
conserved one has to take into account the spectral flow process which
bring the energy levels (together with electrons occupying them) from
one Dirac point to another through the bulk of the valence band, as
schematically shown in Fig.~\ref{fig:1}. Such a possibility exists
only in the presence of a magnetic field.

The existence of the chiral anomaly results in a rather unusual mechanism
for negative magnetoresistance.
The easiest way to calculate the magnitude of the effect is to estimate
the rate of  entropy production in the presence of electric field.
\begin{equation}\label{entropy}
\dot{S}= \sum_{i}\int\! \frac{\drm\bf p}{(2\pi\hbar)^3}\,
  \frac{\bigl(\delta n_{\bf p}^{(i)}\bigr)^2}\tau
  \frac1{n_\p^0(1-n_\p^0)}= \frac{\sigma E^2}T\,.
\end{equation}

At small $\mathbf{E}$, the stationary solution to Eq.~(\ref{kin-eps}) is
\begin{equation}\label{distrpeturb}
  \delta n^{(i)}(\epsilon) = - \frac{k^{(i)}}{\rho^{(i)}(\epsilon)}
  \frac{e^2\tau}{4\pi^2\hbar^2c}
   (\mathbf{E}\cdot\mathbf{B}) \frac{\d n_0(\epsilon)}{\d \epsilon}\,.
\end{equation}
For simplicity let us assume there are only two valleys,
with $k_{1,2}=\pm 1$, and
with the same quasiparticle velocity $v$. Let the $z$-axis be
parallel to ${\bf B}$.
Then, from Eqs.~(\ref{entropy}) and (\ref{distrpeturb}) we get an
anomaly-related contribution to the component $\sigma_{zz}$ of the
conductivity tensor
\begin{equation}\label{sigma}
\sigma_{zz}
 =  \frac{e^2}{4\pi^2\hbar c} \frac vc \frac{(eB)^{2}v^{2}}{\mu^2} \tau.
\end{equation}
Note that $\sigma_{zz}$ given by Eq.~(\ref{sigma}) is an increasing
function of the magnetic field.
All other anomaly-related components of the conductivity tensor
$\sigma_{ij}$ are zero.
In other words, the anomaly-related current can flow only in the direction
of ${\bf B}$.
One can also understand this fact by
 noticing that, at ${\bf E}=0$,  and in the presence of a magnetic field,
Eq.~(\ref{current}) gives
an expression for a current density~\cite{Vilenkin,Fukushima:2008xe}
\begin{equation}\label{current1}
  \mathbf{j} =e\sum_{i} {\bf j}_{i}=
  \frac{e^2}{4\pi^2\hbar^2c} \mathbf{B}  \sum_{i} k^{(i)} \mu^{(i)}.
\end{equation}
Here we assume electron distribution functions in the individual valleys
have equilibrium forms.
In the case of a global equilibrium, all $\mu_{i}=\mu$, and the contributions
to Eq.~(\ref{current1}) from different valleys cancel each other.
According to Eq.~(\ref{chiralnonconserv}), in the presence of electric and
magnetic fields, an imbalance of electron populations and, consequently,
a difference between the
the chemical potentials $\mu_{i}$ is created.  As a result, there is a
finite current density, which  can relax only via intra-valley scattering.
In agreement with Eq.~(\ref{sigma}), its value is proportional to $\tau$,
its direction is parallel to ${\bf B}$, and it responds only on the
component of the electric field parallel to ${\bf B}$.

There is a significant difference between the anomaly-related
[Eq.~(\ref{sigma})] and the conventional Drude contributions
$\sigma^{\textrm{(D)}}_{ij}({\bf B})$ to the  ${\bf B}$-dependence of the
conductivity  tensor. For an isotropic Fermi surface and in the relaxation
time approximation, all components of $\sigma^{\textrm{(D)}}_{ij}$, except for
$\sigma^{\textrm{(D)}}_{zz}$, are decreasing functions of $B$. For
anisotropic Fermi surface, at $(\omega_{\rm c} \tau_{\textrm{tr}})^{2}\ll 1$,
there is
a ${\bf B}$-dependence of  $\sigma^{(D)}_{zz}$ as well, which can be
estimated as
\begin{equation}
 \sigma^{\textrm{(D)}}_{zz}(0)-\sigma^{\textrm{(D)}}_{zz}({\bf B})
  \sim \sigma^{\textrm{(D)}}_{zz}(0)(\omega_{\rm c} \tau_{\textrm{tr}})^{2}.
 \end{equation}
Here $\sigma^{\textrm{(D)}}(0)=e^{2}\nu v^{2}\tau_{\textrm{tr}}/3$ is the
Drude conductivity,
and $\nu\sim \mu^{2}/v^{3}$ is the density of states at the Fermi level.
Thus, for small magnetic field, the anomaly-related contribution
(Eq.~\ref{sigma}) dominates the magnetoresistance, provided
\begin{equation}\label{criterion}
\frac{\tau}{\tau_{\textrm{tr}}}\frac{1}{(\mu \tau_{\textrm{tr}})^{2}} > 1.
\end{equation}
Generically, in small gap semiconductors, the parameter
$\tau/\tau_{\textrm{tr}}\gg 1$, because the inter-valley scattering
requires a large momentum transfer. If the scattering potential is smooth,
this parameter become exponentially large.
Even in the case of an anisotropic Fermi surface, depending on symmetry
there could be a direction of ${\bf E}$ for which $\sigma^{\textrm{(D)}}_{zz}$
is independent of ${\bf B}$.
At small values of
$\mu\ll T$ the conductivity is determined by electron-hole
scattering~\cite{Balcond,Sid}. In this case one has to substitute 
$\mu$ for $T$ in Eqs.~(\ref{sigma}) and (\ref{criterion}), while the parameter
$\tau/\tau_{\textrm{tr}}\gg 1$ is exponentially big. 

For $(\omega_{\rm c} \tau_{\textrm{tr}})^{2}\gg 1$ the ${\bf B}$-dependence of
$\sigma^{\textrm{(D)}}_{zz}({\bf B})$ saturates and it becomes independent
of ${\bf B}$.
In contrast, the deviation  of the anomaly-related contribution from the
quadratic in ${\bf B}$ behavior takes place at much higher magnetic fields.
Thus, $\sigma_{zz}$ could be a non-monotonic function of ${\bf B}$. Finally,
the anomaly-related contribution to the conductivity tensor may be
distinguished by its unusual frequency dependence: it is controlled by
the parameter $(\omega \tau)^{2}$, rather than by the conventional
parameter $(\omega\tau_{\textrm{tr}})^{2}$. Here $\omega$ is the
frequency of the electric field.

At  low values of $\mu$  the anomaly-related contribution to the conductivity
can be even bigger than the Drude contribution $\sigma^{\textrm{(D)}}$.  
In this case system supports a new type of weakly damped
plasma waves with a frequency
\begin{equation}\label{plasma1}
\omega_{\rm p} \sim \pm \sqrt{ \frac{e^2}{\pi \hbar c} \frac vc } \frac{eBv}{T},
\qquad \mu =0
\end{equation}
provided $\omega_{\rm p}\gg \tau^{-1}$.

\begin{figure}[ptb]
\includegraphics[width=0.45\textwidth]{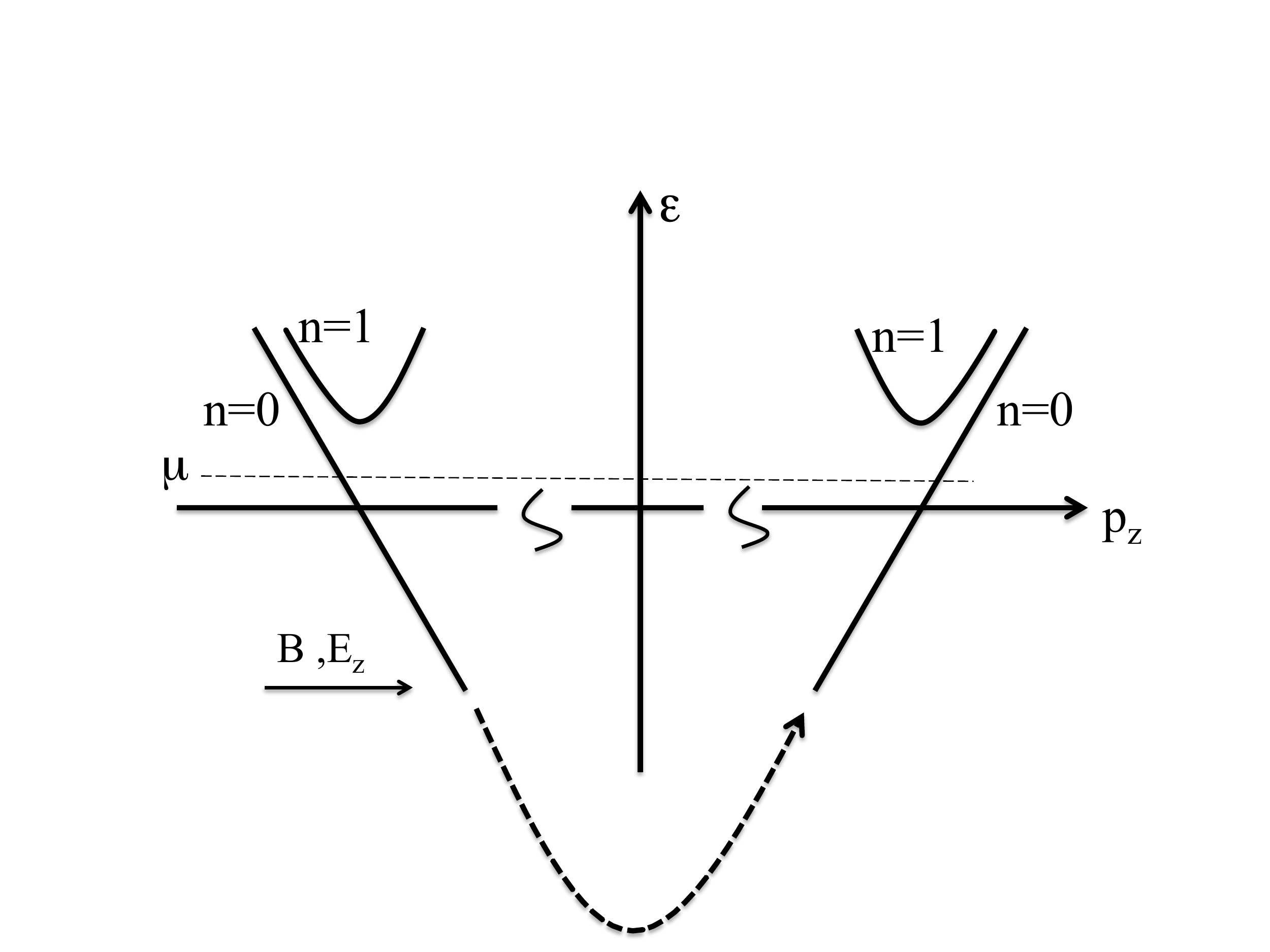}
\caption{Schematic electron spectrum of a Weyl metal in the ultra-quantum limit.
$n=0,1$ label Landau levels. The dashed line indicates the direction of
the electron spectral flow in $p_{z}$ space in the presence a $z$-component
of the electric field.} \label{fig:2}
\end{figure}

The approach based on the semiclassical equations of motion,
Eqs.~(\ref{kineq}) and (\ref{eqmotion}), is valid if
$\mu\gg \hbar \omega_{\rm c}$.
In the opposite, ultra-quantum, limit $\omega_{\rm c}\tau_{\textrm{tr}}\gg 1$,
the anomaly-related negative magnetoresistance has been previously
discussed in Refs.~\cite{Nielsen,Aji}.
In this case the spectrum of the Dirac equation has the form
\begin{equation}
\epsilon_{n}(p_{z})= \pm v\sqrt{ 2n\frac{\hbar e}cB+p^2_{z}} \,, \qquad
  n=1,2, \ldots
\end{equation}
For $n = 0$ case  $\epsilon_{0} = \pm vp_{z}$, where $\pm$ corresponds
to different valleys. In other words, the $n = 0$ Landau level is chiral:
the branches of the spectrum with $\epsilon_{0} = \pm vp_z$ correspond to
different valleys, as shown in Fig.~\ref{fig:2}.
Consider the case where both the chemical potential and the temperature
are small compared to the energy difference between the zero and the first
Landau levels, i.e.  $\mu, T< \hbar v/L_{B}$, where
$L_{B}=\sqrt{\hbar c/eB}$ is the magnetic length. In this case only chiral
branches of the spectrum are occupied by electrons.
Contributions to the current from branches of the spectrum with different
chiralities can relax only by inter-valley scattering processes
characterized by  $\tau$. If the electric field is applied in the $z$
direction, electrons movie according the law
$\dot{p}_{z}=eE_{z}-p_{z}/\tau$; $ v_{z}=\pm v$,
and we get the following expression for the conductivity~\cite{Nielsen}:
\begin{equation}\label{condQuant}
  \sigma_{zz}=\frac{\tau e^{2}v}{4\pi^2 \hbar L_{B}^{2}}.
\end{equation}
By the same token we can obtain an expression  for the plasma frequency
at zero wave vector,
\begin{equation}\label{plazma}
\omega^{2}_{\rm p}=r_{\!s}\frac{2v^{2}}{\pi L_{B}^{2}}\,
\end{equation}
where  $r_{\!s}=e^{2}/\kappa \hbar v$,  and $\kappa$ is the dielectric
constant. Eq.~(\ref{plazma}) is valid if $r_{\!s}<1$.


The fact that Eq.~(\ref{plazma}) does not have a classical limit
($\hbar \rightarrow 0$) is a particular example of a general property of
collective modes in the massless Dirac plasma~\cite{DasSarma}.
We note however, that in three dimensions and at ${\bf B}=0$ the plasma
frequency is proportional to $\mu$.
In contrast, Eq.~(\ref{plazma}) is independent of the value of $\mu$, and
remains finite even when $\mu=0$.

Equations~(\ref{condQuant}) and (\ref{plazma}) are valid only for Weyl
metals where time reversal symmetry is preserved. In systems with no time
reversal symmetry linear in ${\bf B}$ contributions to $\sigma_{zz}$ are
allowed.  The magnitude of these contributions is not universal and
depends on the details  of the mechanism of  time-reversal symmetry
violation.

The work of DTS was supported, in part, by DOE grant No.\ DE-FG02-00ER41132.
The work of BS was supported by NSF grant
DMR-0804151.

\end{document}